\documentclass[journal]{IEEEtran}
\usepackage[T1]{fontenc} 
\usepackage{cite}
\usepackage{algpseudocode}
\usepackage{subfigure}
\usepackage{graphicx}
\usepackage{amsmath}
\usepackage{mathtools}
\usepackage{textcomp}
\usepackage{etoolbox}
\usepackage{comment}
\usepackage{enumitem}
\usepackage{booktabs}
\usepackage{multirow} 
\usepackage{epstopdf}
\usepackage{bm}

\usepackage{gensymb}
\usepackage{array}
\usepackage{wrapfig}
\usepackage{algorithm}
\usepackage[flushleft]{threeparttable}
\usepackage{xcolor}
\usepackage{soul}

\newcommand{\innermid}{\nonscript\;\delimsize\vert\nonscript\;}
\newcommand{\activatebar}{
  \begingroup\lccode`\~=`\|
  \lowercase{\endgroup\let~}\innermid 
  \mathcode`|=\string"8000
}
\usepackage{amsthm,amssymb}
\usepackage{setspace}
\usepackage{rotating,booktabs}
\newcommand{\subparagraph}{}

\usepackage[compact]{titlesec}
\usepackage{hyperref}
\hypersetup{
    colorlinks=false,
}
\usepackage[colorinlistoftodos,prependcaption,textsize=tiny]{todonotes}
\usepackage{bm}
\usepackage{tikz}
\usetikzlibrary{shapes,arrows}

\begin{document}
\title{Foveation-based Deep Video Compression without Motion Search}
\author{Meixu Chen,~\IEEEmembership{Student Member,~IEEE,}
         Richard Webb, and Alan C. Bovik,~\IEEEmembership{Fellow,~IEEE}% <-this % stops a space
\thanks{M. Chen and A. C. Bovik are with the Department of Electrical and Computer Engineering, University of Texas at Austin, Austin, USA (e-mail: chenmx@utexas.edu; bovik@ece.utexas.edu).}
\thanks{Richard webb is with Meta Reality Labs(email: rwebb@fb.com). }}
\maketitle

% As a general rule, do not put math, special symbols or citations in the abstract or keywords.
\begin{abstract}
Virtual Reality (VR) and its applications have attracted significant and increasing attention. However, the requirements of much larger file sizes, different storage formats, and immersive viewing conditions pose significant challenges to the goals of acquiring, transmitting, compressing, and displaying high-quality VR content. At the same time, the great potential of deep learning to advance progress on the video compression problem has driven a significant research effort. Because of the high bandwidth requirements of VR, there has also been significant interest in the use of space-variant, foveated compression protocols. We have integrated these techniques to create an end-to-end deep learning video compression framework. A feature of our new compression model is that it dispenses with the need for expensive search-based motion prediction computations. This is accomplished by exploiting statistical regularities inherent in video motion expressed by displaced frame differences. Foveation protocols are desirable since, unlike traditional flat-panel displays, only a small portion of a video viewed in VR may be visible as a user gazes in any given direction. Moreover, even within a current field of view (FOV), the resolution of retinal neurons rapidly decreases with distance (eccentricity) from the projected point of gaze. In our learning based approach, we implement foveation by introducing a Foveation Generator Unit (FGU) that generates foveation masks which direct the allocation of bits, significantly increasing compression efficiency while making it possible to retain an impression of little to no additional visual loss given an appropriate viewing geometry. Our experiment results reveal that our new compression model, which we call the Foveated MOtionless VIdeo Codec (Foveated MOVI-Codec), is able to efficiently compress videos without computing motion, while outperforming foveated version of both H.264 and H.265 on the widely used UVG dataset and on the HEVC Standard Class B Test Sequences. The Foveated MOVI-Codec project page can be found at \url{https://github.com/Meixu-Chen/Foveated-MOVI-Codec}.

\end{abstract}

% Note that keywords are not normally used for peerreview papers.

\IEEEpeerreviewmaketitle

%Outline:
% 1. VR 的大图片问题
% 2. deep learning的兴起
% 3.HVS特点
% 4.Foveated Video metrics
% 5.我们用的deeplearning和hVS，指标用的什么foveated metric
\section{Introduction}
\IEEEPARstart{R}{ecent} advances in Virtual Reality (VR) offered more immersive viewing experiences than even high-resolution flat-panel displays. However, VR presentations require much larger file sizes, high bandwidths, different storage formats, and spatial-purpose immersive viewing hardware, all of which pose significant challenges against the goals of acquiring, transmitting, compressing, and displaying high quality VR content. One advantage of VR, however, is that the two eyes have fixed positions, aside from eye movements, relative to the viewing screen. Because of this, the eye movements, and associated points of gaze on the displays can be measured. This makes it possible to exploit the fact that the density of retinal photosensors is highly non-uniform. The cone cells used in photopic viewing achieve on peak density in the foveal region, which captures a circumscribed  FOV of about 2.5$\degree$ around gaze. This includes only 0.8\% of all pixels on a flat panel display when viewed under typical conditions \cite{guenter2012foveated}, and around 4\% of pixels on a VR display \cite{patney2016towards, chen2019study}. Since the density of photoreceptors falls away quite rapidly with increased eccentricity relative the fovea, much more efficient representations of what is perceived can be obtained by judiciously removing redundant information from peripheral regions. 

While foveal processing protocols might be useful for many aspects of VR rendering and viewing, such as enhancement or brightening around the point of gaze, foveated compression may offer the most significant and obvious benefits. While this topic has been studied in the past \cite{geisler1998real,wang2001embedded,wang2003foveation}, only recently has there been renewed interest in foveating modern codecs \cite{kaplanyan2019deepfovea}. In this direction, because of their tremendous ability to learn efficient visual representations, deep learning models are viewed as promising vehicles for developing alternative video codecs. This also raises the possibility of creating very efficient, end-to-end deep foveated video compression engines, which is the main topic we study here. Much less work has been done on deep video compression than on learning-based image compression. However, a variety of uniform resolution (without foveation) deep video compression models have been devised \cite{wu2018video,cheng2019learning,yang2020learning,lu2019dvc}. For example, Wu \textit{et al.} \cite{wu2018video} proposed a deep video compression network utilizing the idea that video compression may be conceptualized as image interpolation. Chen \textit{et al.} \cite{cheng2019learning} enhanced spatial-temporal energy compaction in a learning-based video compression model by introducing a spatial energy compaction penalty into the loss function. The authors of \cite{yang2020learning} used bidirectional reference frames to compute motion maps in a hierarchical learning scheme. Lu \textit{et al.} \cite{lu2019dvc} developed a video compression network called DVC, where each component of a traditional hybrid codec is replaced by a deep network. In \cite{chen2022learning}, a deep video compression without motion estimation and compensation is proposed, and the resulting codec is competitive to latest H.266 on high resolution videos against MS-SSIM.

In both traditional video codecs and recent deep learning-based models, motion estimation and compensation occupy a significant portion of compression system resources.  Motion estimation requires expensive search processes that are amplified in very large format videos (e.g. 6K and above) that are needed to display naturalistic video in VR. In the model we proposed here, we avoid motion estimation via search, by instead training a specially-designed network that is able to efficiently represent the residuals between each frame and a set of spatially-displaced neighboring frames. Computing a set of frame differences, even over many displacement directions, is much less expensive than conducting matching-based search processes. Moreover, while the statistics of motion vectors are generally not regular, the intrinsic statistics of frame differences exhibit strong regularities\cite{soundararajan2012video}, especially when the differences are taken between spatially displaced lying along the same local motion paths \cite{lee2021space}. The strong internal structure of aligned, high-correlation frame differences makes them sparser and easier to efficiently represent in a deep architecture.

Our idea is inspired by the way the human visual system processes natural time-varying images. Many studies have produced strong evidence suggesting that the early stages of vision are primarily implicated in reducing redundancies in the sensed input visual signals \cite{atick1990towards,attneave1954some}. Indeed, much of early visual processing along the retino-cortical pathway appears to be devoted to processes of spatial and temporal decorrelation\cite{dong1995temporal, rucci2015unsteady, chichilnisky2002functional, engbert2006microsaccades, poletti2016compact, olshausen1996emergence}. We have found that sets of spatially displaced frame differences, which are space-time processes, supply a rich and general way to exploit space-time redundancies\cite{soundararajan2012video,lee2021space}. Our idea is also related to recent theories of the role of microsaccades in human visual information processing\cite{rucci2015unsteady, chichilnisky2002functional, engbert2006microsaccades, poletti2016compact}. Microsaccades create small spatial displacements of the visual field from moment to moment. While microsaccades have been theorized to play roles in avoiding retinal saturation, maintaining accurate fixation in the presence of drifts, and preserving the perception of fine spatial details \cite{poletti2016compact}, it has been more recently theorized to play an important role in efficiently representing locally changing and shifting space-time visual information \cite{rucci2015unsteady, chichilnisky2002functional, engbert2006microsaccades, poletti2016compact}. We believe that micro-saccadic eye movements may be an efficient adaptation to efficiently exploit local regularities induced by small spatial displacements over time, to achieve more efficient visual (neural) representations. This has inspired us to, in like manner, train a deep foveated coder-decoder network that compresses videos using regular displaced residual representations as inputs.

Because the compute bandwidths and data volumes involved in VR rendering are also unusually high, we have sought to reduce both data and computation in two ways: by perceptually relevant foveation, and by a perception-driven elimination of expensive motion computations.

The success of foveation based processing protocols involves several factors, including distribution of retinal ganglion cells \cite{purves2001functional}, cortical magnification \cite{harvey2011relationship}, and the steep grade of density of the photoreceptors \cite{wassle1990retinal}. The spacings of the photoreceptors and the receptive ﬁelds of the neurons they feed are smallest in the fovea \cite{cheung2016emergence}. The fovea covers an area in the approximate range of 0.8\% to 4\% of the pixels on a display, depending on the display size, resolution, and the assumed typical viewing distance \cite{guenter2012foveated, patney2016towards}. Recent advances in eye-tracking technology and their integration into consumer VR headsets have opened the possibility of using them to facilitate gaze-contingent video compression. Indeed, retinal foveation when combined with ballistic saccadic eye movements to direct visual resources, is a form of biological information compression. For example, the density of retinal ganglion cells (RGC) in the fovea is 325,000/$mm^2$. If the entire retina had this output density, then about 350 million RGCs would be implied. However, the number of axons carrying signals along the optic nerves of each eye is only around 1 million, hence foveation results in a 350-fold compression of data passed along the retino-cortical pathway \cite{weber2009implementations}. In an analogous manner, considerable increases in digital video compression can be obtained by removing visual redundancies (relative to fixation) in the visual periphery. 

Here we introduce a foveated deep video compression network that is efficient, statistically and perceptually motivated, and free to motion search computations. Our specific contributions may be summarized as follows:
\begin{itemize}
    \item We incorporate foveation into a deep video compression model to achieve significant data reductions suitable for eye-tracked VR systems.
    \item We innovate the use of displaced frame differences to capture efficient representations of structures and temporal statistical redundancies induced by motion.
    \item The overall video compression system, which we call the Foveated MOVI-Codec is optimized using a single loss function.
    \item The new model is shown to obtain state-of-the-art compression performance on the widely used UVG dataset and on the HEVC Standard Test Sequence Class B dataset, outperforming H.264, H.265 and their foveated counterparts against the well-known perceptual video quality index Foveated Wavelet Image Quality Index(FWQI)\cite{wang2001embedded,wang2001foveated}.
\end{itemize}

The rest of the paper is organized as follows. Section \ref{section2} details the architecture and training protocol used to create the Foveated MOVI-Codec model. Section \ref{section3} explains the experiments on algorithm performance and comparisons that we conducted. Section \ref{section4} concludes the paper with a discussion of future research directions. 

\section{Related Works}
\subsection{Foveated Video Compression}
Since the turn of the millennium, there has been a slowly growing interest in the use of foveation for such diverse image and video processing tasks as quality assessment \cite{lee2002foveated}, segmentation \cite{boccignone2005foveated}, and watermarking \cite{koz2002foveated}. Methods of foveating visual content can be categorized into three ways: geometric transformations, space-varying filters, and space-variant multiresolution decompositions \cite{wang2017foveated}. In the first of these, a foveated retinal sampling geometry is used to either apply a foveating coordinate transformation on an original uniform resolution image \cite{wang2001rate}, or to average and map local pixel groups into superpixels \cite{wallace1994space, tsumura1996image}. Filter-based methods process images with space-varying low-pass filter with cut-off frequencies determined by foveated resolution-reduction protocols \cite{liu2005foveation, sheikh2002foveated}. Multiresolution methods foveation involves decomposing images into bandpass scales, and only retaining scales specified by a foveal fall-off function defined relative to a measured or presumed fixation point \cite{burt1988smart, geisler1998real}.

Recently, given significant advances in high resolution and immersive displays technologies, along with concurrent increases in VR content, interest of foveation as an efficient processing tool has quickened. Recent related models include \cite{li2011visual}, where a neurobiological model of visual attention is used to predict high saliency regions and to generate saliency maps. A guidance map is also generated, using foveation to guide bit allocations when tuning quantization parameters in video compression system. Li \textit{et al.} \cite{li2018learning} trained a content-weighted CNN to conduct image compression, whereby the bitrates allocated to different parts of an image are adapted to the local content. Their system significantly outperforms JPEG and JEPG2000 in terms of SSIM when operating in a low bitrate regime. Mentzer \textit{et al.} \cite{mentzer2018conditional} proposed a similar but simpler model, by incorporating a second channel at the output of the encoder that is expanded into a mask which is used to modify the latent representations. DeepFovea \cite{kaplanyan2019deepfovea} is a foveated reconstruction model, that employs a generative adversarial neural network. A peripheral video is reconstructed from a small fraction of pixels, by finding a closest matching video to the sparse input stream of pixels that lies on the learned manifold of natural videos. This method is fast enough to drive gaze-contingent head-mounted displays in real time.

\subsection{Foveated Video Quality Assessment}
When designing foveated compression systems, it is desirable to be able to access their perceptual efficiencies using quality measurement tools that account for the foveation. However, almost all available image quality measurement tools, such as SSIM \cite{wang2004image}, operate on spatially uniform resolution contents. However, there are a few foveated video quality assessment models, which can be conveniently divided into several types. One type of foveated VQA model uses purely static, spatial foveation, whereby measurement or prediction of the user's point of gaze guides the space variant measurement of quality as a function of eccentricity. For example, the Foveated Wavelet Image Quality Index (FWQI) utilizes wavelets to extract position-dependent spatial quality information \cite{wang2001foveated, wang2001embedded}. Several factors are taken into consideration, including the spatial contrast sensitivity function, which is used to determine local visual cutoff frequencies, which guides modeling of human visual sensitivity across the available wavelet subbands, when combined with assumption on viewing distance and the display resolution. Lee \textit{et al.} \cite{lee2002foveated} proposed a foveal signal-to-noise ratio (FSNR) to evaluate the quality of picture or video streams. In this method, a foveated image is obtained by a foveated coordinate transformation on the original image(s) to be quality-accessed.

A second type of foveated VQA model is based on retinal velocity. In addition to static foveation mechanisms, these kinds of models also take advantage of the fact that the contrast sensitivity of HVS to an object in a moving scene is influenced by the velocity of its map on the retina. Movement in a video may cause two effects: loss of acuity of the moving objects, modifications of perceived quality. Further, two factors can contribute to losses of acuity: increases of retinal image velocity, and increases of eccentricity relative to the foveal center. Based on these observations, Riomac-Drlje \textit{et al.} \cite{rimac2010foveated} proposed a foveated mean squared error (FMSE) that models the effects of spatial acuity reduction due to motion. Another model called the foveation-based content Adaptive Structural SIMilarity index (FA-SSIM), which is based on the popular IQA model SSIM \cite{rimac2011foveation} combines SSIM with a foveation-based sensitivity function. 

You \textit{et al.} \cite{you2014attention} proposed a full reference attention-driven foveated video quality metric (AFViQ) that accounts for the localization of fixations in images and videos. All of the algorithms mentioned above assume that the point of fixation is the center of the image, which is not always true, and can lead to an invalid foveation model. As a result, algorithms based on automatic fixation detection have also been proposed. AFViQ attempted to solve this problem by integrating foveation into a wavelet-based distortion visibility model.

\section{Proposed Method}
\label{section2}
\subsection{Framework}
\begin{figure*} [tbp]
\centerline{
\includegraphics[width=1.3\columnwidth]{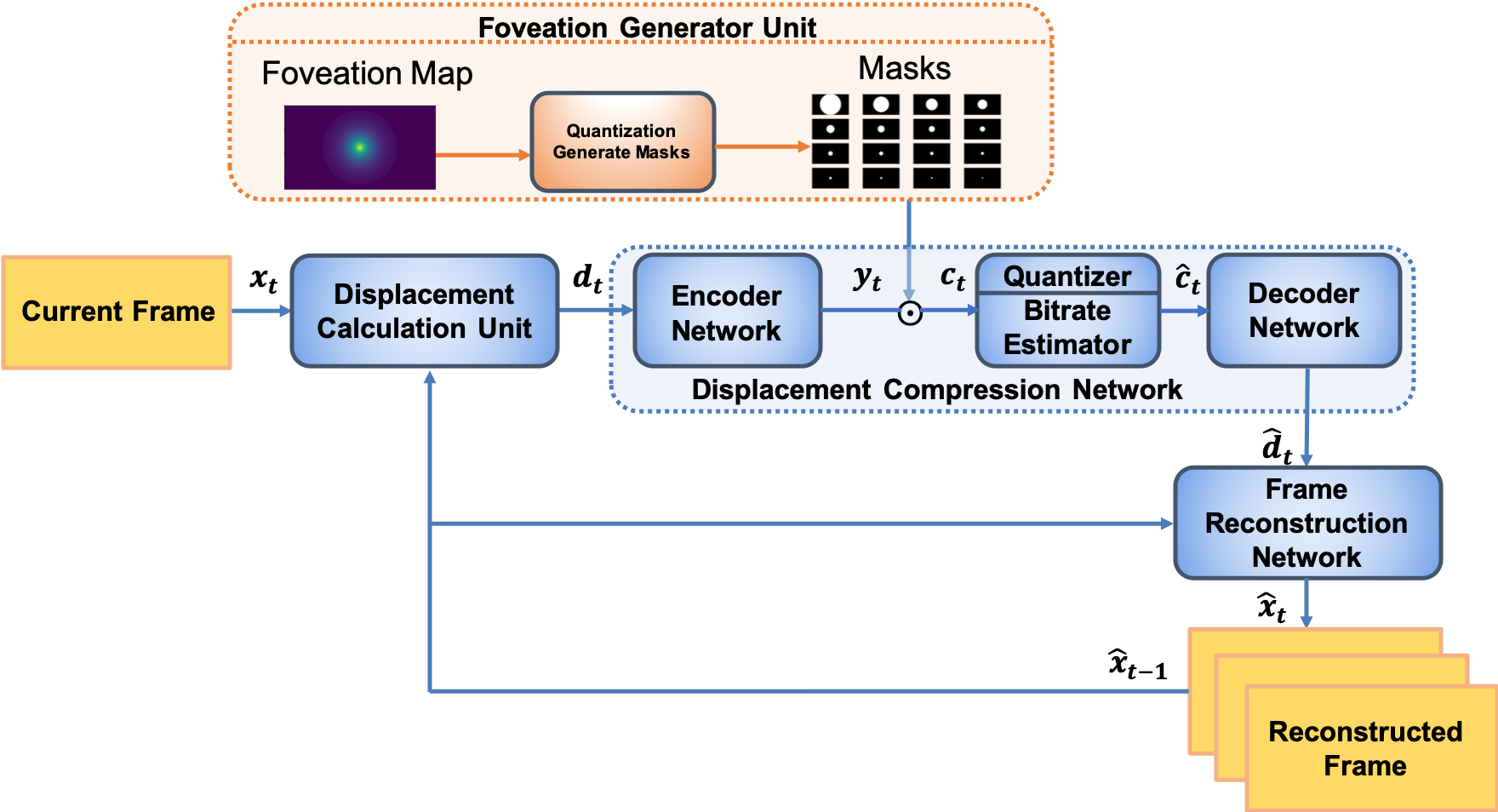}
}
\caption{Overall network architecture of the Foveated MOVI-Codec, which consists of four components: a Displacement Calculation Unit, a Displacement Compression Network, a Foveation Generation Unit, and a Frame Reconstruction Network.} 
\label{fig:network}
\end{figure*}

Figure \ref{fig:network} illustrates the overall architecture of our network, which extends our previous MOVI-Codec \cite{chen2022learning}. The compression network is comprised of four components: a Displacement Calculation Unit (DCU), a Displacement Compression Network (DCN), a Foveation Generator Unit (FGU), and a Frame Reconstruction Network (FRN). The DCU computes displaced frame differences between the current frame and the previous reconstructed frame; the FGU generates foveation masks that later direct the allocation of bits in DCN; the DCN compresses displaced frame differences generated from DCU; and the FRN reconstructs input frames from the previous reconstructed frame and the reconstructed displaced frame differences. 

The flow of our network is: Given an input video with frames $x_1, x_2, ..., x_T$, for every frame $x_t$, calculated displaced frame differences between the current frame $x_t$ and previous reconstructed frame $\hat{x}_{t-1}$ via the DCU, after which the displaced frame differences $d_t$ are input into the DCN. In the FGU, a perception-based foveation map $P$ is generated from \cite{geisler1998real, wang2001foveated} and used to generate a set of foveation masks $M(P)$. After the set of displaced frame differences $d_t$ are encoded into latent representations $y_t$, the masks generated from the FGU direct the allocation of bits via element-wise multiplication of $y_t$ and $M(P)$, producing a masked latent representation $c_t$, which is then quantized (via rounding) and decoded to $\hat{d_t}$. Finally, the FRN reconstructs the input frame $\hat{x_t}$ from the reconstructed displaced frame differences $\hat{d_t}$ and the previous reconstructed frames $\hat{x}_{t-1}$. The DCN and FRN are defined identically as in \cite{chen2020learning}, so we do not further elaborate them here. We explain the DCU and FGU in the following.

\subsection{Displacement Calculation Unit (DCU)}
\begin{figure} [!t]
\centerline{
\includegraphics[width=0.8\columnwidth]{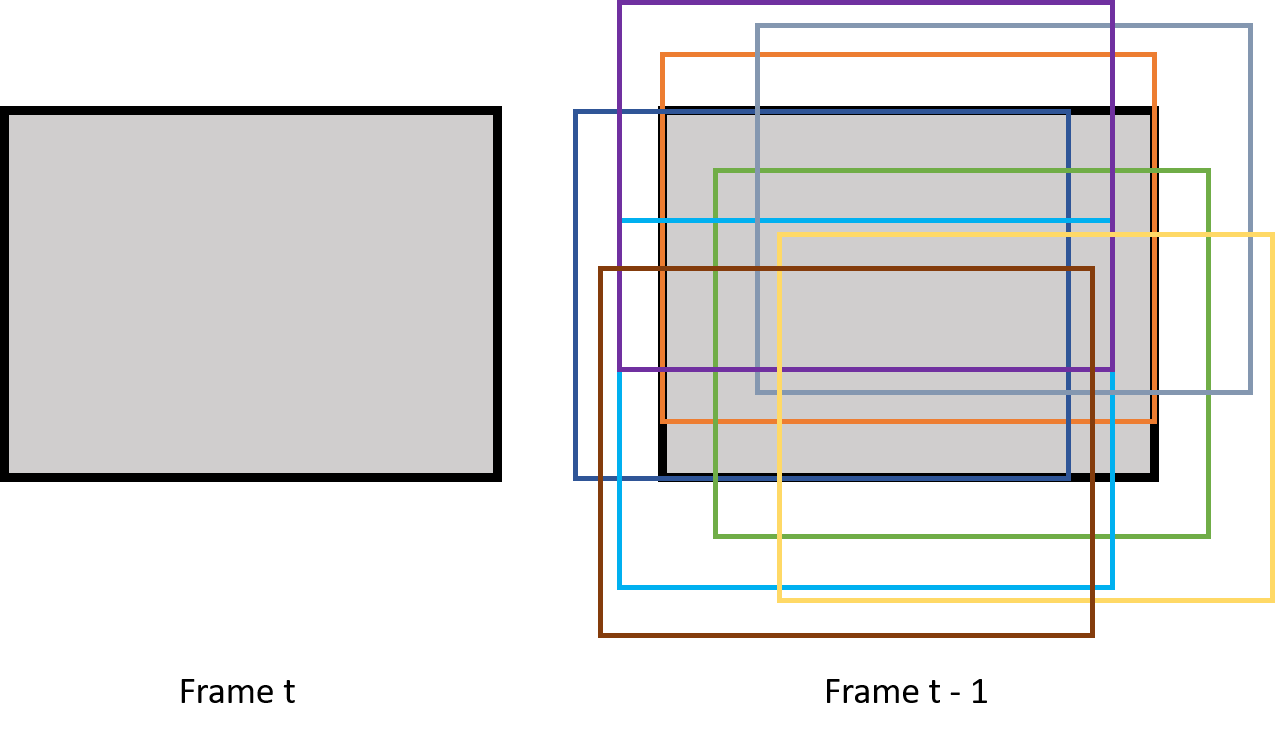}
}
\caption{Concept of displaced frame differences, showing a frame $t$ and a previous frame $t - 1 $, and multiple spatially displaced versions of frame $t - 1$ that can also be differenced with frame $t$.}
\label{fig:displacement}
\end{figure}

The DCU removes the need for any kind of motion vector search. Instead, the DCU learns to optimally represent time-varying images as sets of spatially displaced frame differences. Given a video with $T$ frames ${x_1, x_2, ..., x_T}$ of width $W$ and height $H$, two directional (spatially displaced) temporal differences are computed between each pair of adjacent frames, as shown in Figure \ref{fig:displacement}. Assume that the inputs to the DCU a current frame $x_t$ and a reconstructed previous frame $\hat{x}_{t-1}$. Then, at each spatial coordinate $(i, j)$, a set of spatially displaced differences is calculated as: 
    \begin{equation}
        d_H(i,j)_{t} = x_t(i,j) - \hat{x}_{t-1}(i, j - s),
    \end{equation}
    \begin{equation}
        d_V(i,j)_{t} = x_t(i,j) - \hat{x}_{t-1}(i - s, j).
    \end{equation}
In our current implementation, $s = 0, \pm{3}, \pm{5}, \pm{7}$. This set of 13 displaced frame differences (residuals) is then fed into the DCU, which delivers as output the reconstructed set of displaced residuals $\hat{d_t}$. As mentioned in Section \ref{section2}, the statistics of non-displaced frame differences have been observed to be highly regular \cite{soundararajan2012video}. As shown in \cite{lee2021space}, the statistics of displaced frame differences are also regular, and more so in the direction of local motion. This makes them good video representations to learn to exploit space-time redundancies, while avoiding the computational burdens of search-based motion estimation and compensation. Although the range of motions between frames can be larger than our largest choice of displacement, larger motions can be captured by combinations of our set of displacements.

\subsection{Foveation Generator Unit (FGU)}
\label{section:FGU}
In the DCN, the encoded video data that is output from the quantizer is still spatially invariant, and arithmetic coding is used to further compress the code. However, the goal of the FGU is to exploit the non-uniform distribution of ganglion cells and photoreceptors across the visual field. Our basic tool to accomplish this is an established model of the contrast sensitivity function (CSF) expressed in terms of eccentricity. We use this to enable increased compression of the image in a manner such that the reconstructed frames are indistinguishable from the original around the point of fixation, as well as with increasing eccentricity. 

A good model of the contrast threshold is given by \cite{geisler1998real}:
\begin{equation}
    CT(f,e) = CT_0 exp(\alpha f \frac{e + e_2}{e_2}),
\end{equation}
where $f$ is spatial frequency, $e$ is retinal eccentricity, $CT_0$ is a specialized minimum contrast threshold, $\alpha$  is a spatial frequency decay constant and $e_2$ is the half-resolution eccentricity. We follow best fitting parameter values given in \cite{geisler1998real} are $\alpha = 0.106$, $e2 = 2.3$, and $CT0 = 1/64$ in our experiment. The CSF is then:
\begin{equation}
    CS(f,e) = \frac{1}{CT(f,e)}.
\end{equation}
The authors of \cite{wang2001foveated} defined a foveation-based error sensitivity in terms of viewing distance $D$, frequency $f$, and location $(x, y)$:
\begin{equation}
\label{eq:foveation}
\begin{gathered}
    S_f(D, f, x,y) = \\
    \begin{cases}
    \frac{CS(f, e(D, x, y))}{CS(f,0)} = exp(-\alpha f \frac{e(D,x,y)}{e_2}) & \text{for } f \leq f_m(x)\\
    0 & \text{for } f > f_m(x)
    \end{cases},
\end{gathered}
\end{equation}
where $f_m$ is the cutoff frequency.

\begin{figure} [htp]
\centering
\subfigure{
  \label{img}
\includegraphics[width=0.4\columnwidth]{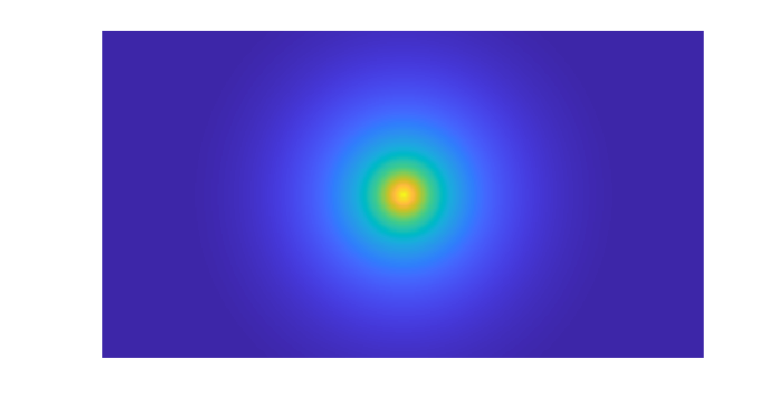}
\includegraphics[width=0.4\columnwidth]{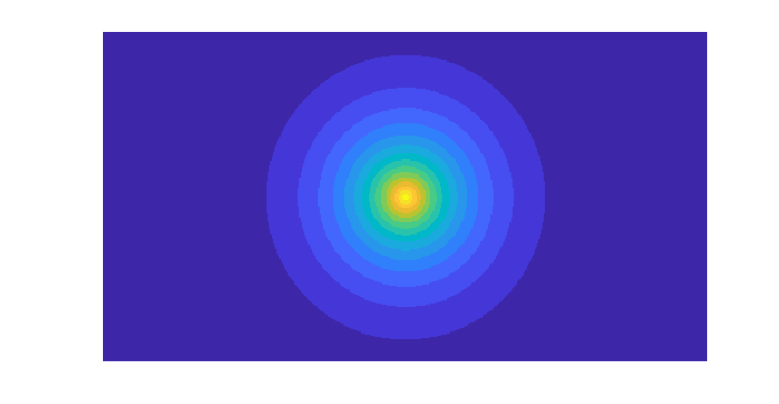}}
\caption{Foveation map (left) and quantized foveation map (right), where brighter regions corresponds to larger value.}
\label{fig:foveation_map}
\end{figure}

\begin{figure} [htp]
\centering
\subfigure{
  \label{img}
\includegraphics[width=1\columnwidth]{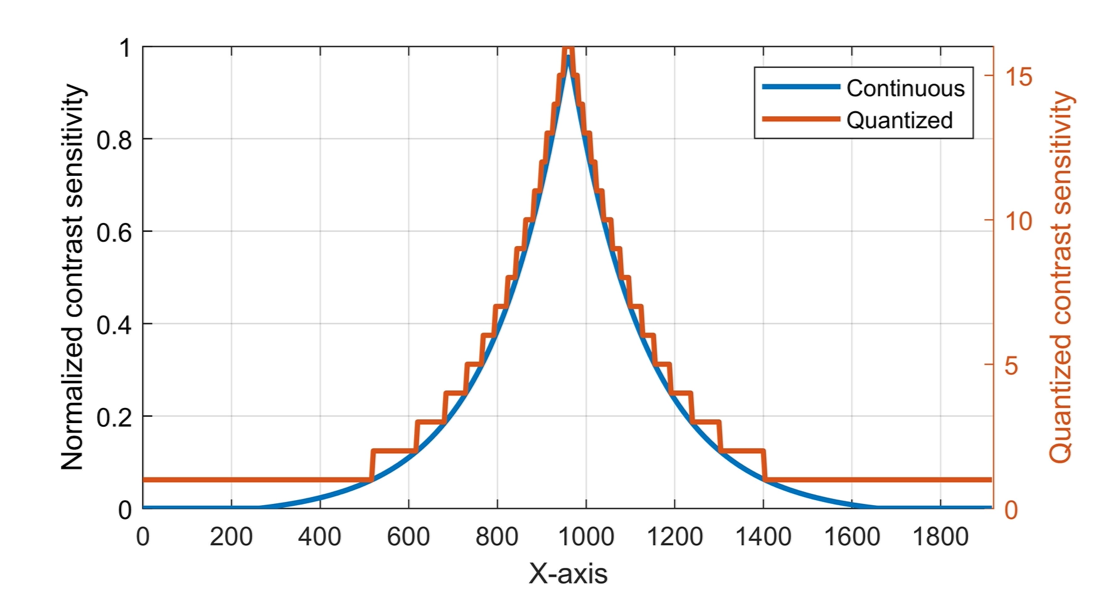}}
\caption{Quantized contrast sensitivity function.}
\label{fig:quant_CSF}
\end{figure}

In our model, we fix the frequency in Equation  \ref{eq:foveation} to be the maximum frequency that can be presented on the display without aliasing. The FGU uses these models to generate a foveation map that is used to guide bit allocation and rate control. During training, foveation maps are generated using Equation \ref{eq:foveation}, assuming a fixed screen resolution, center gaze, and viewing distance following \cite{kaplanyan2019deepfovea} as shown in Figure \ref{fig:foveation_map}. In Equation \ref{eq:foveation}, the contrast sensitivity decays forwards zero beyond the cutoff frequency. Our approach to foveation is quantum; rather than changing the displayed resolutions in a smooth and graded manner, which makes the problem more complex, it is instead quantized. Quantization is applied to yield $n$ levels of the foveation map, and $n=16$ in the current implementation. We also make sure that the contrast sensitivity for the last level is larger than zero to be able to reconstruct all periperal information. Figure \ref{fig:foveation_map} shows a quantized foveation map. Since the latent representations for the displaced frame differences $d_t$ are 128 channels, the same mask is assigned for every 8 channels of latent representations. The quantized map in x axis is shown in Figure \ref{fig:quant_CSF}. After a set of $n$ masks $M(P)$ are generated, we element-wise multiply $M(P)$ and the encoder output $y_t$ to obtain quantized spatially variant (foveated) codes $c_t$ which are then subjected to entropy coding and bitrate estimation, using the same procedure as \cite{chen2022learning}. 

While quantized foveation maps are used to train our model, during application (testing) we instead use isotropic 2D gaussian shaped foveation maps, where the gaussians are defined to follow the modified fall-off of visual acuity. This allows for smoother perceived changes of foveation, with the significant added benefit of making it possible to effect variable rate control by varying the widths ($\sigma$) of the gaussians. We define this parameter as foveation mask space constant (FMSC).

\subsection{Bit Rate Allocation}
Given an input frame $x$, let $y = E(x) \in R^{c \times h \times w}$ be the output of the encoder network, which includes $c$ feature maps of sizes of $h \times w$. Also let $p = P(x)$ denote a $h \times w$ non-negative foveation map to be applied. The expand $y$ using masks $\mathbf{m} \in R^{c \times h \times w}$ as follows:
\begin{equation}
m(i, j, k) = \left\{
        \begin{array}{ll}
            1 & \quad \text{if } p(i, j) \geq \lfloor \frac{k}{c/L} \rfloor \cdot \frac{1}{L}\\
            0 & \quad \text{others}
        \end{array},
    \right.
\end{equation}
where $c$ is the number of channels in the latent representations $y$, and $L$ is the number of desired compression levels across the foveation regions. In this way, more bits are allocated to the foveal region, preserving visual details with less sacrifice of bit rate. The sum of the foveation maps $\sum_{i,j} p_{i,j}$ naturally serves as a continuous estimate of compression rate, and can be directly adopted as a compression rate controller. Because of the flexibility of this foveation map approach, it is not necessary to apply entropy rate estimation when training the encoder and decoder, using a simple binarizer for quantization of latent representations $y$. 

\subsection{Training Strategy}
We are able to model the loss function considering only the distortion as follows:
\begin{equation}
\label{loss}
D = [{D_1(x_t, \hat{x}_t)} + \beta D_2(d_t, \hat{d}_t)] ,
\end{equation}
where $D$ represents the distortion, and $D_1$ is the distortion between the input frame $x_t$ and reconstructed frame $\hat{x}_t$, measured by foveation-weighted SSIM as detailed below at the end of this subsection. $D_2$ is the distortion between the displaced frame differences $d_t$ and the reconstructed displaced frame differences $\hat{d}_t$, as measured by the MSE. The weight $\beta$ controls the trade-off between the perceptual distortion $D_1$ and the pixel-to-pixel distortion $D_2$. 

To leverage multi-frame information in our RNN-based codec structure, we update the network parameters every set of $N$ frames during model training, using the loss function in Equation \ref{loss}, but modified to be a sum of losses over the $k$th set of $N$ frames indexed $x_{{t_k}+1}, ..., x_{{t_k}+N}$:
\begin{equation}
 D_k  = \frac{1}{N}\sum_{n=1}^{N}[{D_1(x_{{t_k}+n}, \hat{x}_{{t_k}+n})} + \beta D_2(d_{{t_k}+n}, \hat{d}_{{t_k}+n})].   
\end{equation}

\begin{figure*} [htp]
\centering
\subfigure{
  \label{img}
\includegraphics[width=1.5\columnwidth]{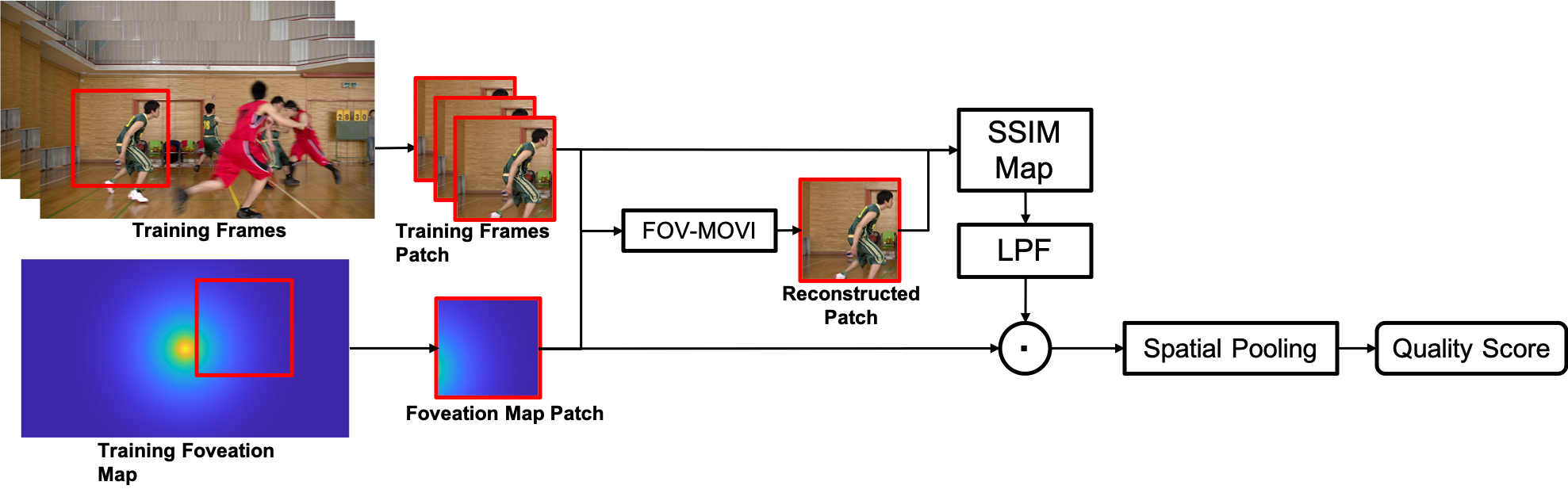}}
\caption{Training strategy.}
\label{fig:training}
\end{figure*}

During training, we selected a random $W \times W$ patch from each training video, and also randomly sampled a patch of the same size from the foveation map, to generate foveation masks from the patch. Foveation-weighted SSIM scores were calculated by applying a low-pass filter (Haar's filter) on the SSIM scores of each frame patch, then multiplying them by the foveation map patchs. The overall workflow is shown in Figure \ref{fig:training}.

\section{Experiments}
\label{section3}
\subsection{Settings}
The Foveated MOVI-Codec networks that we experimented with were trained end-to-end on the  Kinetics-600 dataset\cite{kay2017kinetics,carreira2018short} and on the Vimeo-90K dataset\cite{xue2019video}. We used part of the testing set from Kinetics-600, which consists of around 10,000 videos, to conduct our experiments. From each video, a random 192 $\times$ 192 patch containing 49 frames was randomly selected for training, and normalized the values of each input video to [-1,1]. Since Kinetics-600 dataset consist of YouTube videos of different resolutions, we  randomly downsampled each original frames and extracted a 192 $\times$ 192 patches from the foveation maps to reduce any previously introduced compression artifacts. We randomly sampled 192 $\times$ 192 patches from the foveation maps to generate foveation masks for bitrate allocation. The Vimeo-90K dataset consists of 4,278 videos of fixed resolution 448 $\times$ 256. Since the videos in this dataset each have 7 frames, we randomly selected patches from each of the same size as mentioned earlier (overall 7 frames) for training.  

We fixed the mini-batch size to 8 for training, while the step length $N$ of the recurrent network was set as 7. We used Adamax optimizer for training and set the initial learning rate to 0.0001. The whole system is implemented based on PyTorch and using one Titan RTX GPU. By training on both the Vimeo-90K and the Kinetics-600 datasets, we are able to generalize our model to a wider range of natural motions. We tested the Foveated MOVI-Codec on the JCT-VC Class B datasets \cite{sullivan2012overview} and the UVG datasets \cite{UVG}. Both of these testing datasets have HD resolution contents (1920 $\times$ 1080).

\begin{figure} [!h]
\centerline{
\includegraphics[width=0.8\columnwidth]{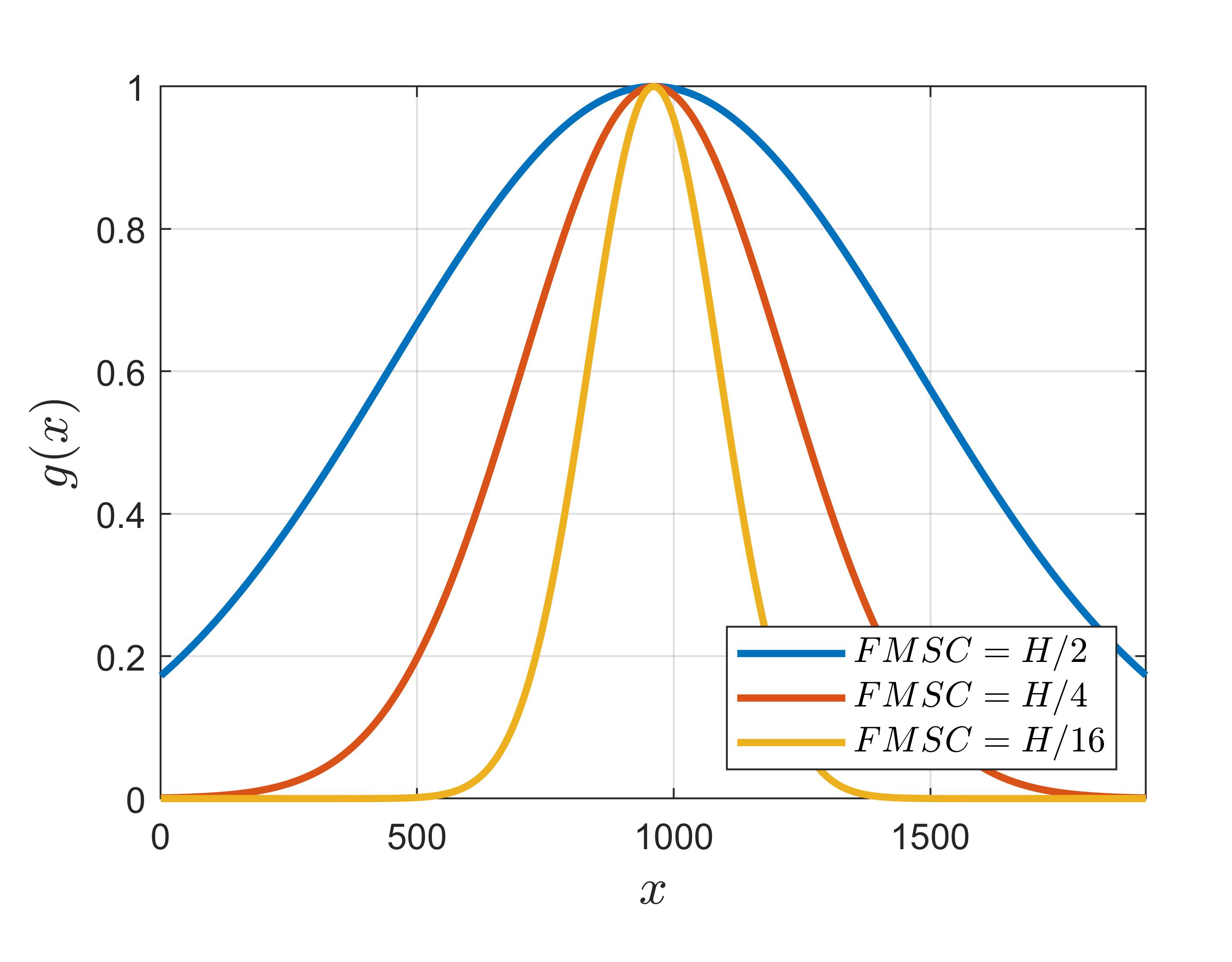}} 
\caption{Normalized sliced profiles of gaussian foveation masks.}
\label{fig:gaussian}
\end{figure}

\begin{figure*} [!h]
\centerline{
\subfigure{
  \label{mask1}
\includegraphics[width=0.5\columnwidth]{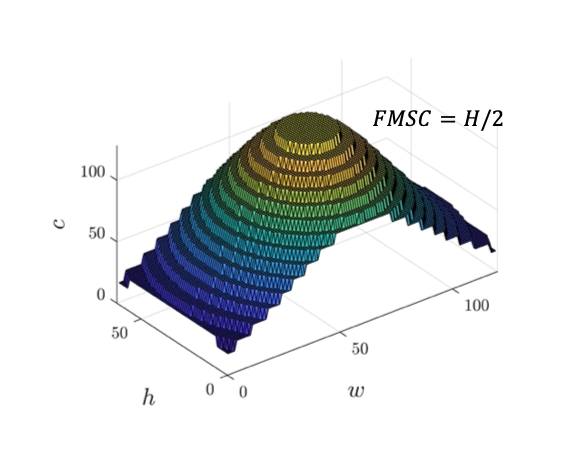}} 
\subfigure{
  \label{mask2}
\includegraphics[width=0.5\columnwidth]{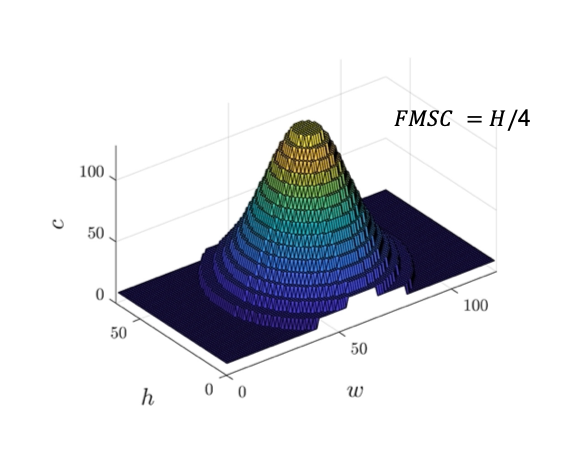}}
\subfigure{
  \label{mask3}
\includegraphics[width=0.5\columnwidth]{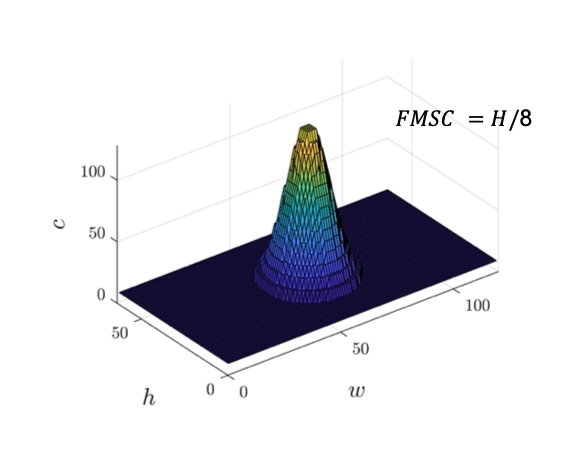}}
}
\caption{Examplar quantized gaussian foveation masks with different foveation mask space constants $FMSC$s.}
\label{fig:masks}
\end{figure*}

In order to assess the reconstruction quality of the foveation compressed videos, we utilized the perceptually relevant FWQI foveated video quality measurement tool follwing the same settings in \cite{kaplanyan2019deepfovea}, with screen width being 0.02 meters and display distance being 0.012 meters. We also used the foveated SSIM model which deploys a fixed foveation map generated from the error sensitivity function from \cite{wang2001foveated}. During testing, videos having different bitrates were generated using gaussian shape foveation maps with different foveation mask space constants, e.g. FMSCs of $\frac{H}{10}$,$\frac{H}{8}$,$\frac{H}{6}$,$\frac{H}{4}$,$\frac{H}{3}$, and $\frac{H}{2}$, where $H$ is the height of the input frame. Examplar 1D slices through the gaussians are shown in Figure \ref{fig:gaussian}, while the corresponding quantized maps are shown in Figure \ref{fig:masks}. We fixed $\beta = 1$.  
\subsection{Results}
\subsubsection{Rate-Distortion Curve}
We compared our video compression engine against the standardized hybrid codecs H.264 and H.265, and also against our previous non-foveated model, the MOVI-Codec, on the UVG dataset and the HEVC Standard Test Sequences Class B. In addition, we also implemented a foveated version of the hybrid codecs using the foveation method mentioned \cite{geisler1998real}. Both testing datasets have resolutions 1920 $\times$ 1080.

\begin{figure} [!h]
\centerline{
\subfigure{
  \label{UVG}
\includegraphics[width=0.5\columnwidth]{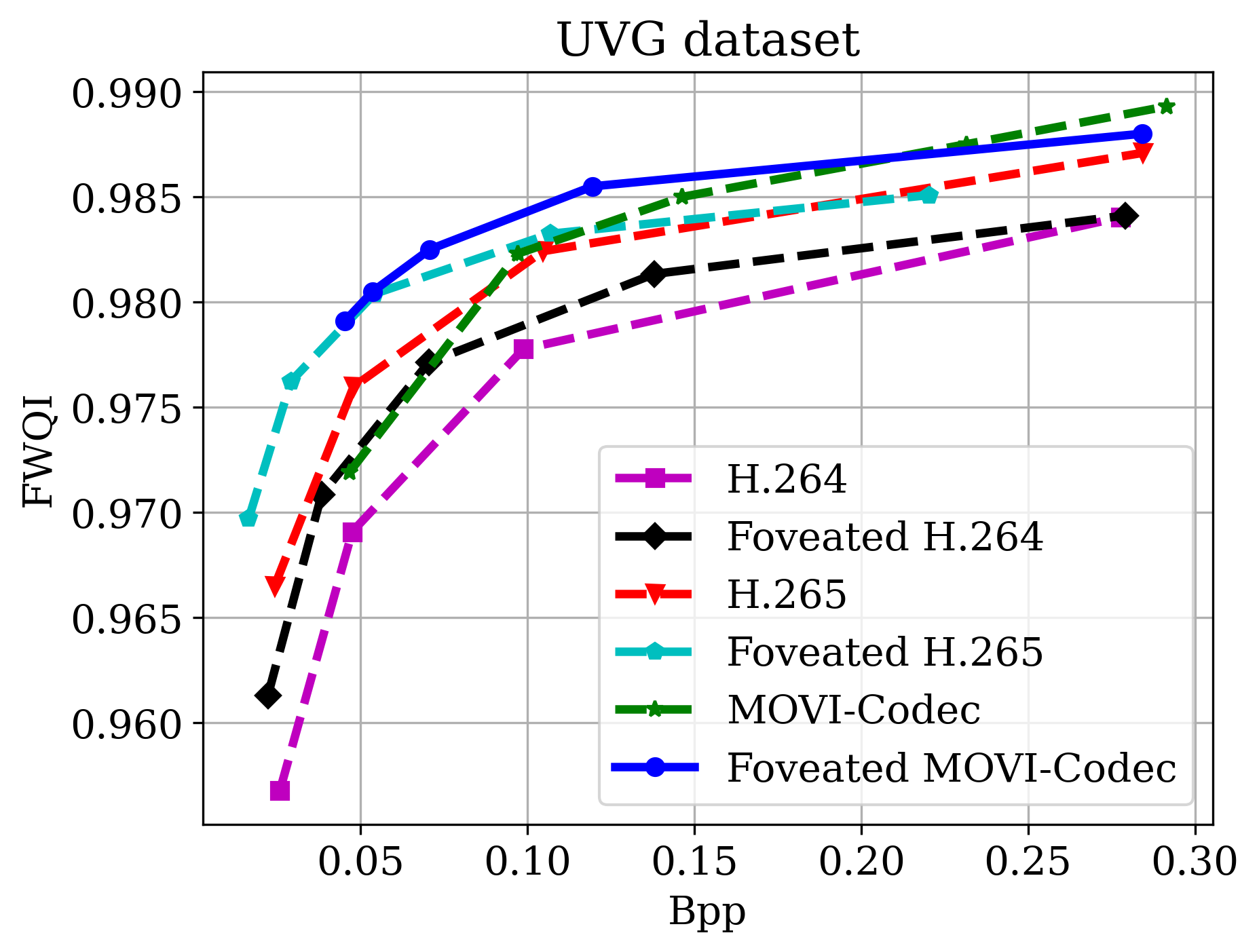}} 
\subfigure{
  \label{HEVC}
\includegraphics[width=0.5\columnwidth]{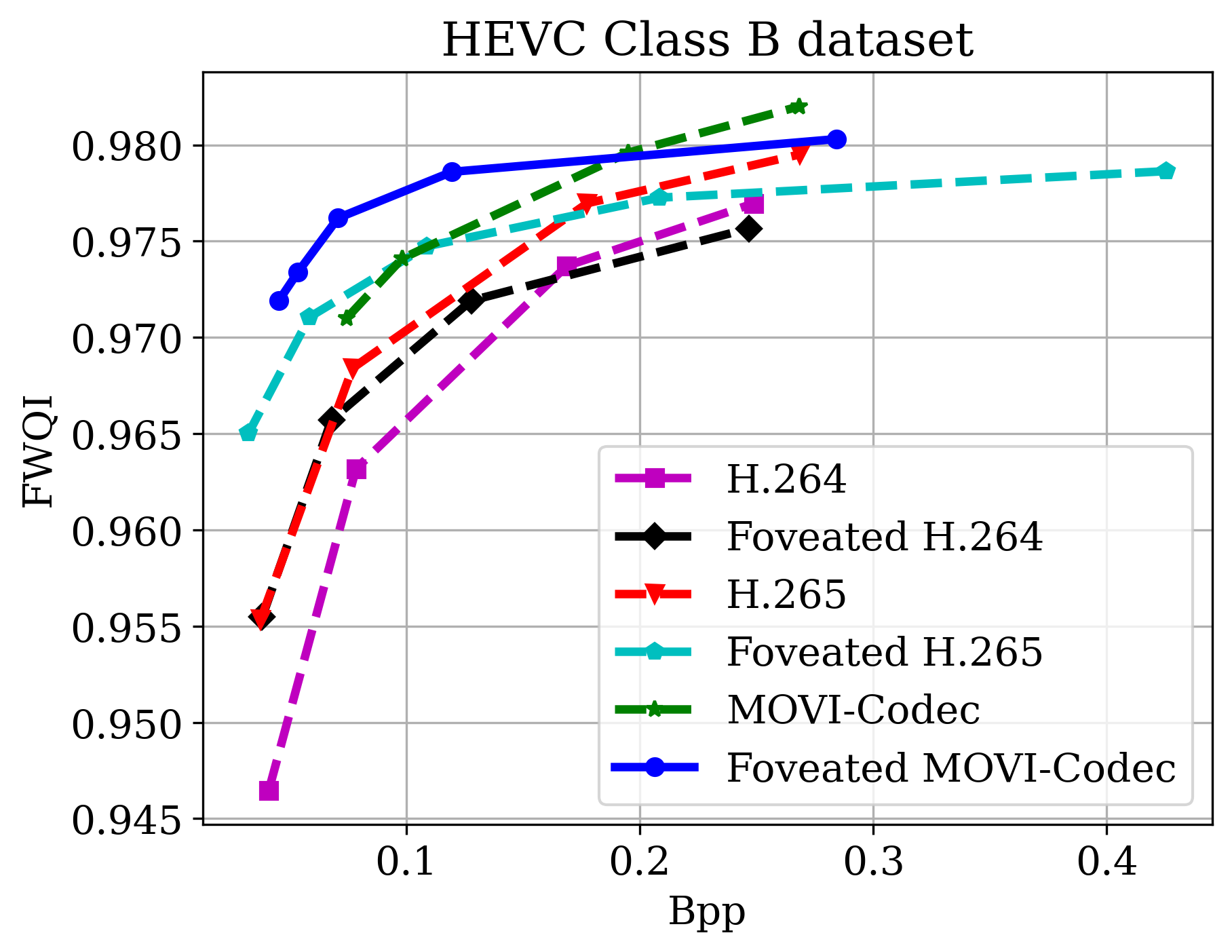}}
}
\caption{FWQI of the compared models on the UVG dataset and HEVC B test sequences. All video resolutions are 1920 $\times$ 1080.}
\label{fig:FWQI}
\end{figure}

\begin{figure*} [htp]
\centering
\subfigure[Basketball Drive]{
  \label{img}
\includegraphics[width=1.8\columnwidth]{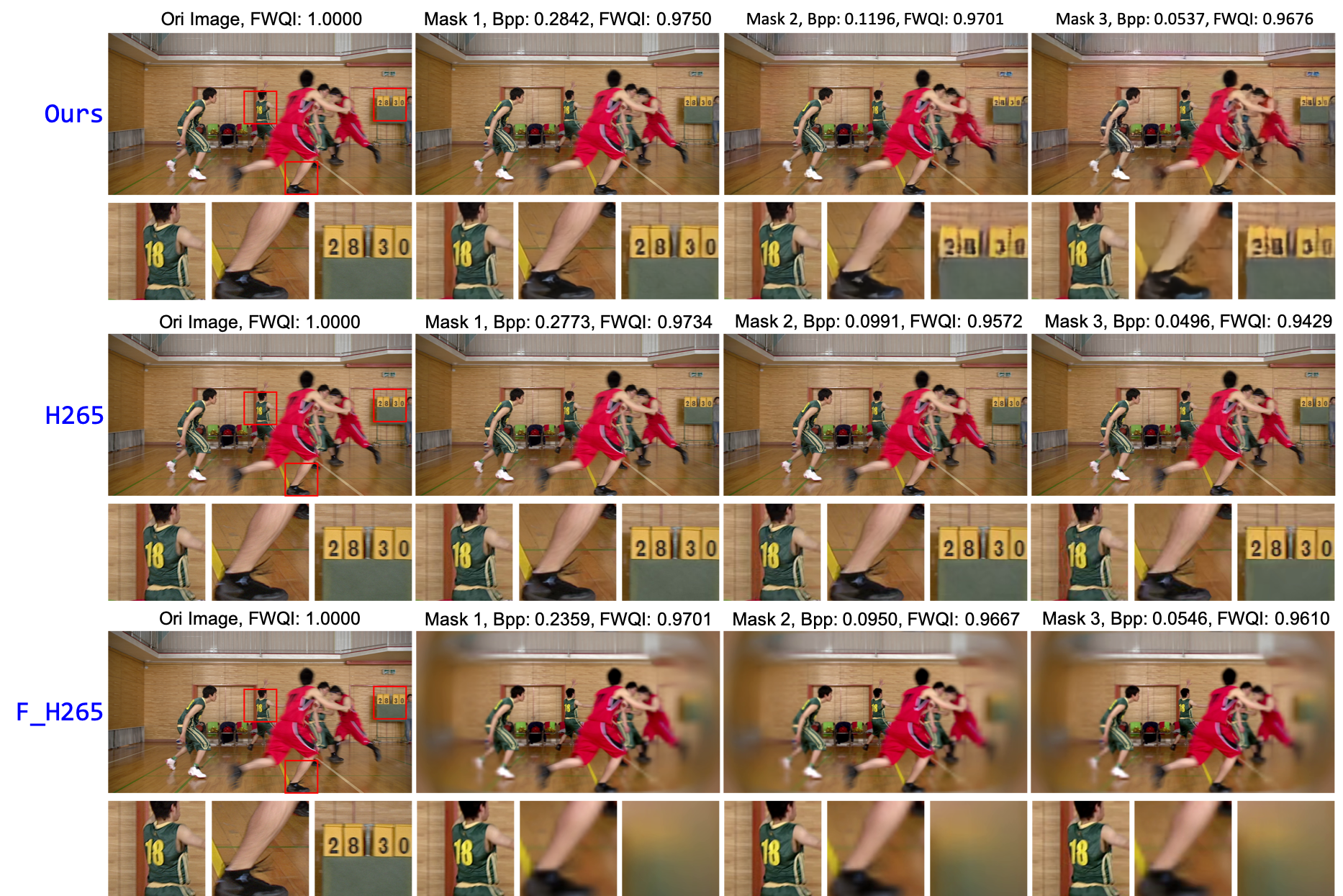}}
% \centering
% \subfigure[Kimono]{
%   \label{img}
% \includegraphics[width=1.5\columnwidth]{Figures/Visuals/Kimono.png}}
\centering
\subfigure[Cactus]{
  \label{img}
\includegraphics[width=1.8\columnwidth]{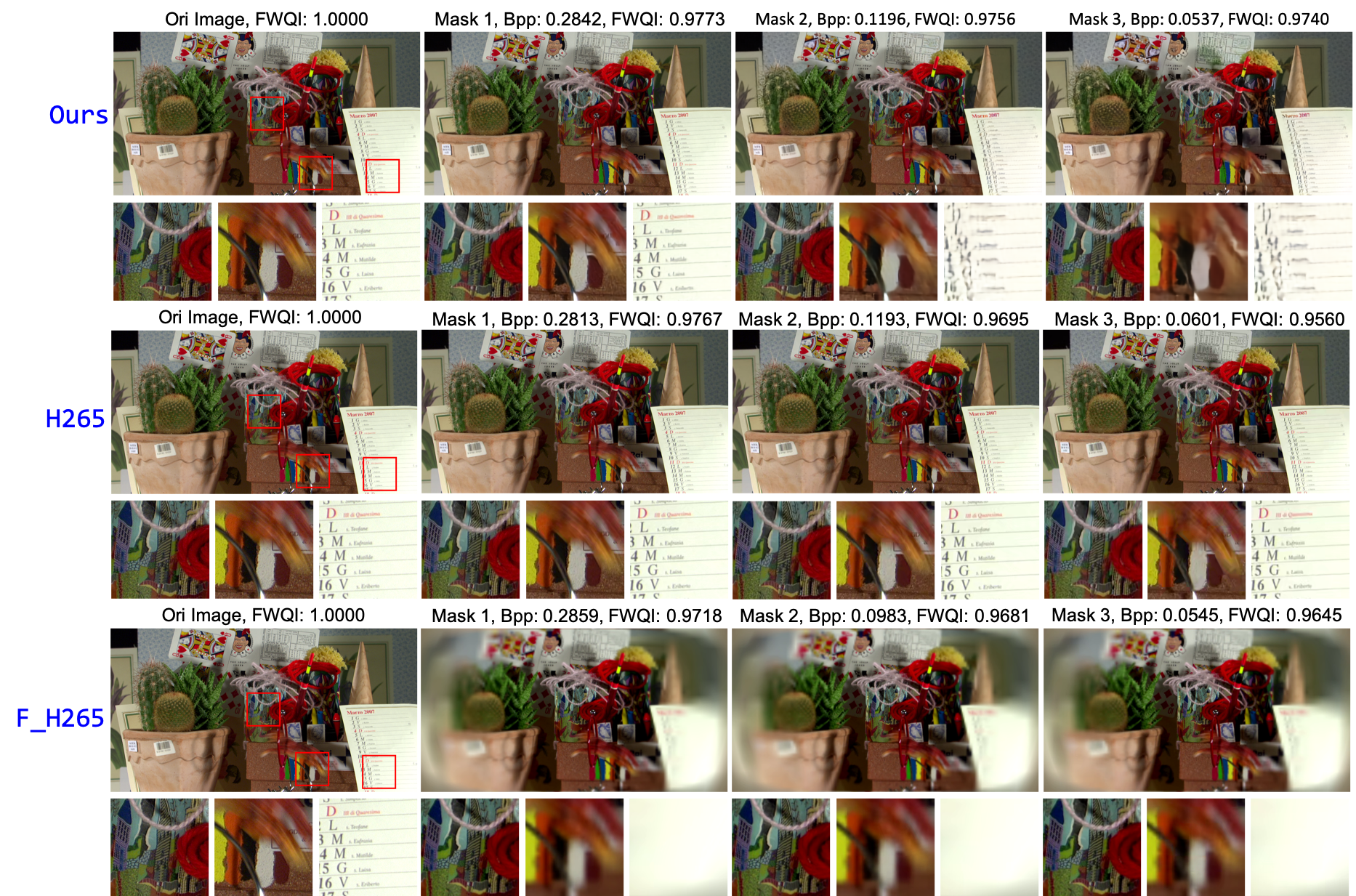}}
\caption{Visualizations of examplar foveated frames reconstructed by FOV-MOVI-Codec, H.265, and Foveated H.265 (denoted F\_265) on the videos (a) Basketball drive and (b) Cactus.}
\label{fig:Visual}
\end{figure*}

Figure \ref{fig:FWQI} shows the results obtained on the UVG and HEVC Class B datasets. Unsurprisingly, the foveated version of the hybrid codecs outperforms their foveated counterparts in terms of FWQI. These results also show that our foveated model outperformed the non-foveated MOVI-Codec on both datasets. Moreover, the Foveated MOVI-Codec outperformed both H.264 and H.265, as well as their foveated counterparts, on both datasets. It is worth noting that the measured qualities of the reconstructed videos produced by Foveated MOVI-Codec produced did not vary much with respect to bitrate, suggested that our model is able to maintain a high quality fovea, while decreasing the bitrate derived from the periphery without sacrificing perceptual video quality. Visualizations of example frames compressed using different levels of bitrates and qualities are shown in Figure \ref{fig:Visual}. More exemplar reconstructed videos are included on our project page with link given in the Abstract. In these reconstructed frames, we selected three regions for detailed comparison: one in the foveal region and the other two others in peripheral. Our model is able to reconstruct videos having higher quality foveas and peripheral regions than the compared models, both visual and in terms of FWQI.


\subsubsection{Latent Representations}
\begin{figure*} [tb]
\centerline{
\includegraphics[width=1.8\columnwidth]{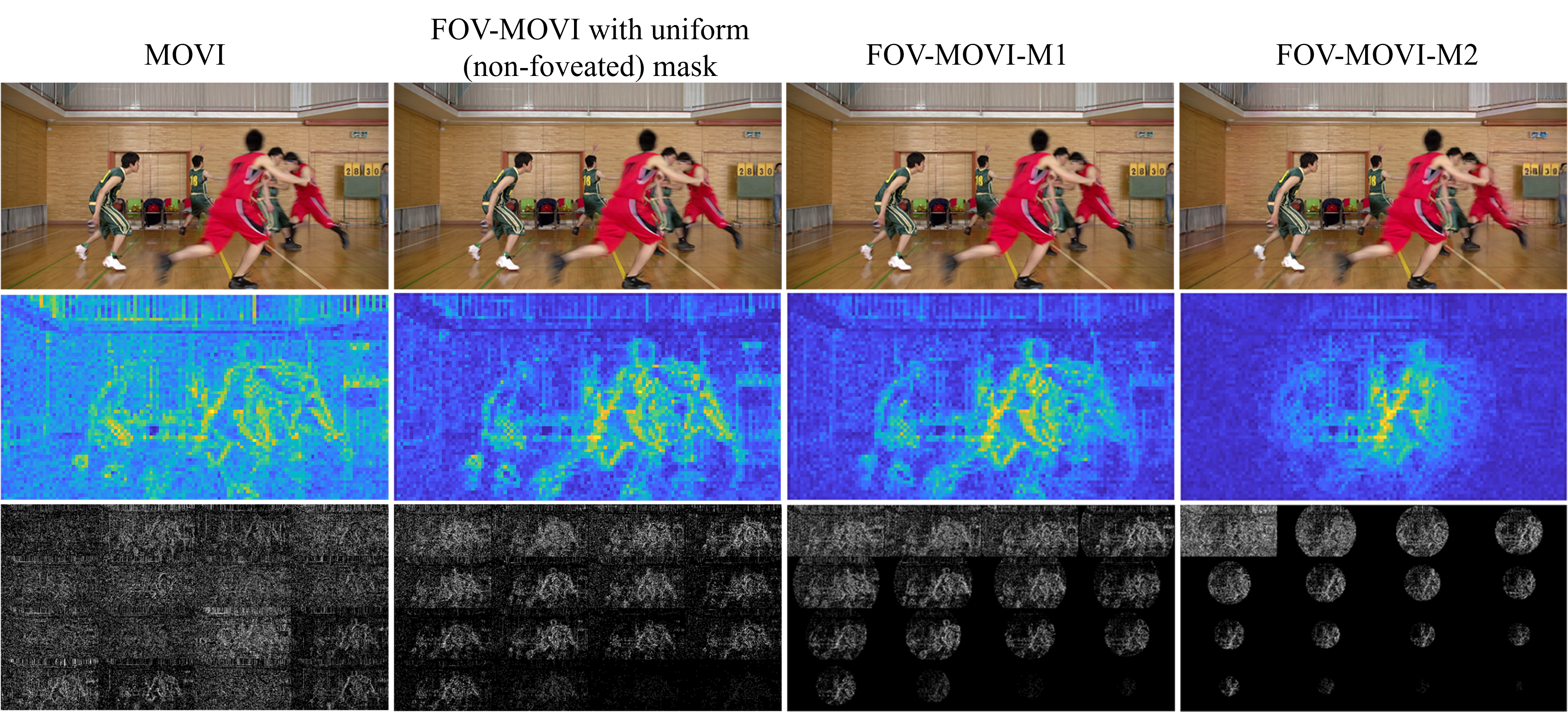}} 
\caption{Latent representations generated from four models. The first row correponds to reconstructed frames from each model, the second row shows the cumulative latent representations, and the last row shows the latent representations at each compression level. FOV-MOVI-M1 is Foveated MOVI-Codec with foveation mask space constant $FMSC = H/2$ and FOV-MOV-M2 is Foveated MOVI-Codec with $FMSC = H/4$, where $H$ is the height of the frame.}
\label{fig:mask_latent}
\end{figure*}

\begin{figure} [tb]
\centerline{
\subfigure[MOVI-Codec]{
  \label{channel}
\includegraphics[width=0.5\columnwidth]{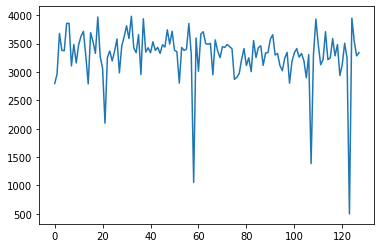}} 
\subfigure[Foveated MOVI-Codec]{
  \label{F_channel}
\includegraphics[width=0.5\columnwidth]{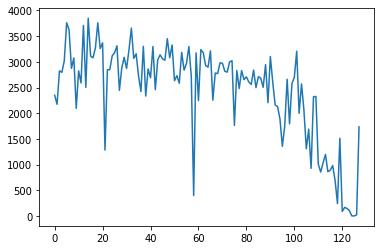}}
}
\caption{Sum of latent representations for each channel, where the sum is decreasing in foveated version.}
\label{fig:channelplot}
\end{figure}

As mentioned in Section \ref{section:FGU}, the Foveated MOVI-Codec uses foveation maps to mediate bit allocations as a function of eccentricity relative to visual fixation. To visualize this process, we compared the latent representations (the encoded outputs) $y_t$ in the Foveated MOVI-Codec against the encoded outputs $y'_t$ of the original MOVI-Codec as shown in Figure \ref{fig:mask_latent}. In the figure, the first row corresponds to reconstructed frames under different models, where the first column shows reconstructed frames from the MOVI-Codec, the second column contains reconstruction from the Foveated MOVI-Codec trained with a uniform (non-foveated) importance map with the masks of first $N$ channels being one and zero elsewhere, and $N$ is a random number during training.  The remaining two columns show reconstructions from the Foveated MOVI-Codec with  foveation mask space constants FMSCs equal to $H/2$ and $H/4$, respectively. The second row shows the corresponding accumulated feature maps, where brighter colors correspond to larger numbers of more features. This shows that more features are used to represent the foveal region, as the foveation maps become narrower (smaller FMSC). The last row of Figure \ref{fig:mask_latent} shows the latent representations at each level (8 channels per level) of the reconstructed frames. Figure \ref{fig:channelplot} also shows the sum of latent representations of $y_t$ and $y'_t$ (foveated and non-foveated, respectively). As shown in Figure \ref{fig:channelplot}, the sum is roughly flat for the non-foveated compressor, whereas the sum decreases as with the channel number for the foveated compressor. This suggests that the foveated network was able to learn more relevant features in the first few channels. From Figures \ref{fig:mask_latent}-\ref{fig:channelplot}, we may conclude that the model learned efficient features across channels and bit allocation, even without the masked multiplication.


\subsubsection{Bit Allocation}
\begin{figure*} [!h]
\centerline{
\subfigure[Reconstructed frame with $FMSC = H/2$]{
  \label{mask1}
\includegraphics[width=0.6\columnwidth]{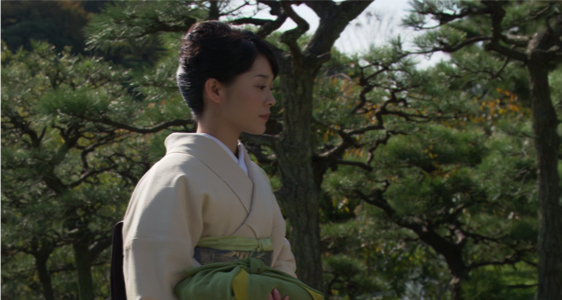}} 
\subfigure[Reconstructed frame with $FMSC = H/4$]{
  \label{mask2}
\includegraphics[width=0.6\columnwidth]{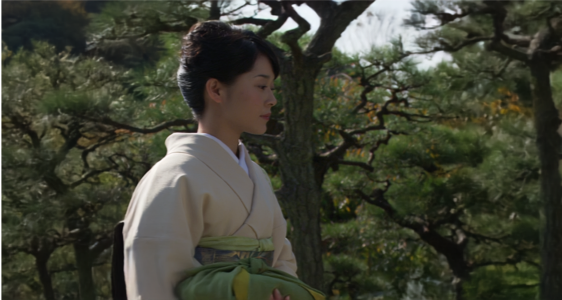}}
\subfigure[Reconstructed frame with $FMSC = H/6$]{
  \label{mask3}
\includegraphics[width=0.6\columnwidth]{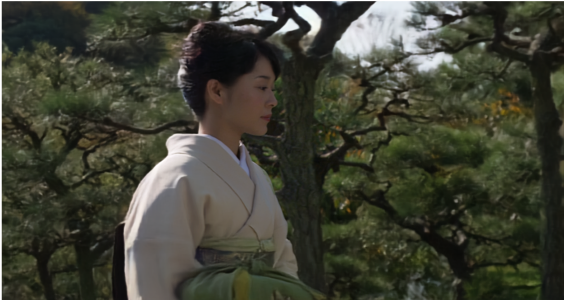}}
}
\centerline{
\subfigure[Differenced frames with $FMSC = H/2$]{
  \label{mask1}
\includegraphics[width=0.6\columnwidth]{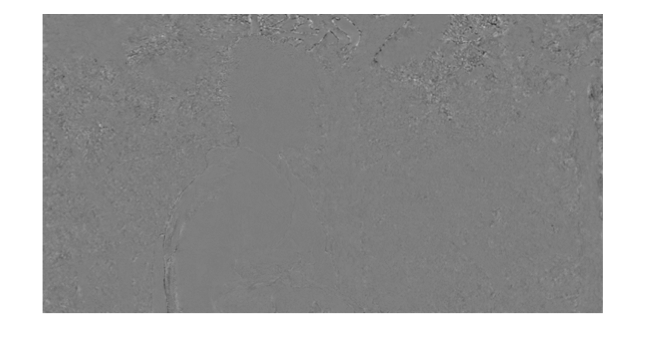}} 
\subfigure[Differenced frames with $FMSC = H/4$]{
  \label{mask2}
\includegraphics[width=0.6\columnwidth]{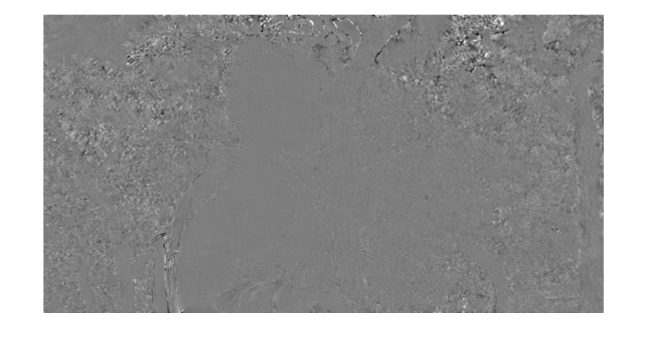}} 
\subfigure[Differenced frames with $FMSC = H/6$]{
  \label{mask3}
\includegraphics[width=0.6\columnwidth]{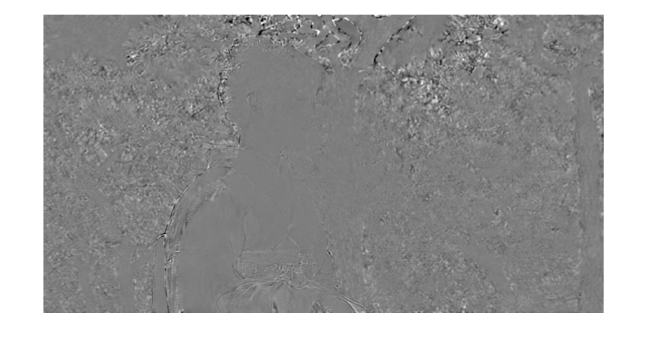}} 
}
\centerline{
\subfigure[Bits and SSIM Profile with $FMSC = H/2$]{
  \label{mask1}
\includegraphics[width=0.6\columnwidth]{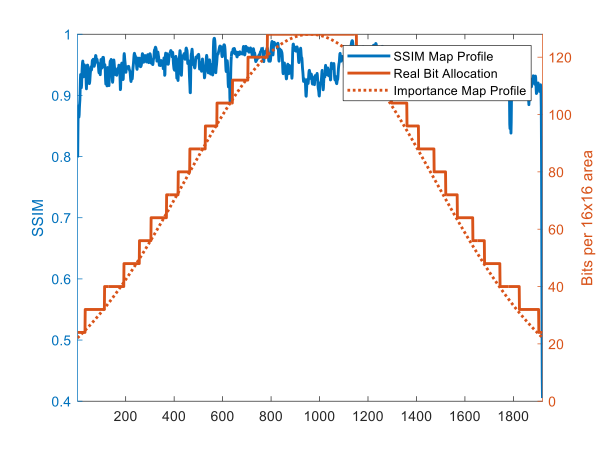}} 
\subfigure[Bits and SSIM Profile with $FMSC = H/4$]{
  \label{mask2}
\includegraphics[width=0.6\columnwidth]{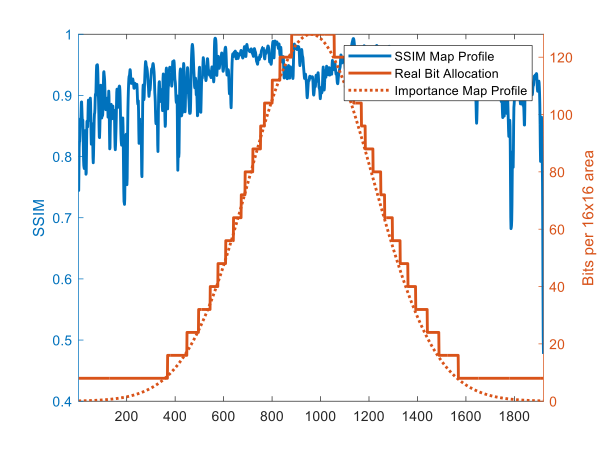}}
\subfigure[Bits and SSIM Profile with $FMSC = H/6$]{
  \label{mask3}
\includegraphics[width=0.6\columnwidth]{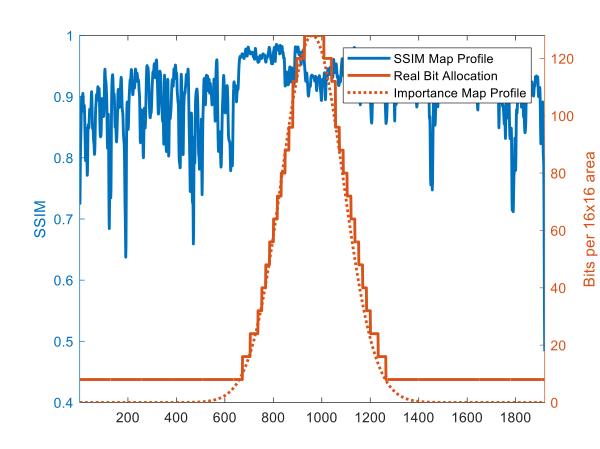}}
}
\caption{Reconstructed frames, differenced frames and bit-SSIM profiles under different foveation space constants (FMSCs).}
\label{fig:bits}
\end{figure*}

Figure \ref{fig:bits} shows the reconstructed frames from the Foveated MOVI-Codec, differenced frames between original frames and reconstructed frames, and bits and SSIM profiles, when using different foveation space constants. From the differenced frames, we can conclude that our model is able to reconstruct a foveal region similar to the original frame regardless of the mask used. The third row of Figure \ref{fig:bits} shows the both a bit allocation plot and the SSIM map profile for different models. From the SSIM map profile, it may be observed that the lower number of bits allocated to the peripheral does not result in lower quality, since the quality of the reconstructed frames remain similar overall.
\subsection{Discussion}
Our experiments have shown that deploying foveation masks leads to much more efficient video compression for suitable environments, such as VR. Our new model outperformed H.264, H.265 and their foveated counterparts against FWQI across all testing sequences. Our model is best targeted at high resolution, gaze contingent foveated compression applications in VR and AR. The hierarchical masks make it possible to transmit scalably, viz., the first levels of content when bandwidth is limited, followed by the other levels. Since foveation masks are used in our model, the first transmitted levels correspond to foveal regions which draw the attention, and are the most important, supplying additional efficiency related to traditional hybrid codecs. Further, the new method is faster than MOVI-Codec since it does not require arithmetic coding. In the current model, the foveation masks are fixed with respect to frame height. One future direction is to train sets of masks adaptive to contents. Another direction is to extend the framework to generate a foveation map based on frequency as well and use it to allocate the contents learnt in the latent representation.

\section{Conclusion}
\label{section4}
We have proposed an end-to-end deep learning video compression framework that assigns bits according to a foveation protocol, assuming known visual fixations. We also achieve efficiency by training a deep space-time compression network to use displaced frame differences to compute efficient motion information by learning optimal between-frame interpolated representations. Our experimental results show that our approach, which we call FOV-MOVI-Codec, outperforms both H.264 and H.265 and foveated versions of them. The low complexity of our model, which avoids motion search, could make it amenable for implementations on resource-limited devices, such as smartphones, VR headsets, and AR glasses.

% \appendices
% \section{Proof of the First Zonklar Equation}
% Appendix one text goes here.

% % you can choose not to have a title for an appendix
% % if you want by leaving the argument blank
% \section{}
% Appendix two text goes here.

% % use section* for acknowledgment
% \section*{Acknowledgment}

% The authors would like to thank...

% Can use something like this to put references on a page
% by themselves when using endfloat and the captionsoff option.
% \ifCLASSOPTIONcaptionsoff
%   \newpage
% \fi

% 
\bibliographystyle{IEEEtran}
\bibliography{IEEEexample}{}
\clearpage
\begin{figure*} [htp]
\centering
\subfigure{
  \label{img}
\includegraphics[width=1.8\columnwidth]{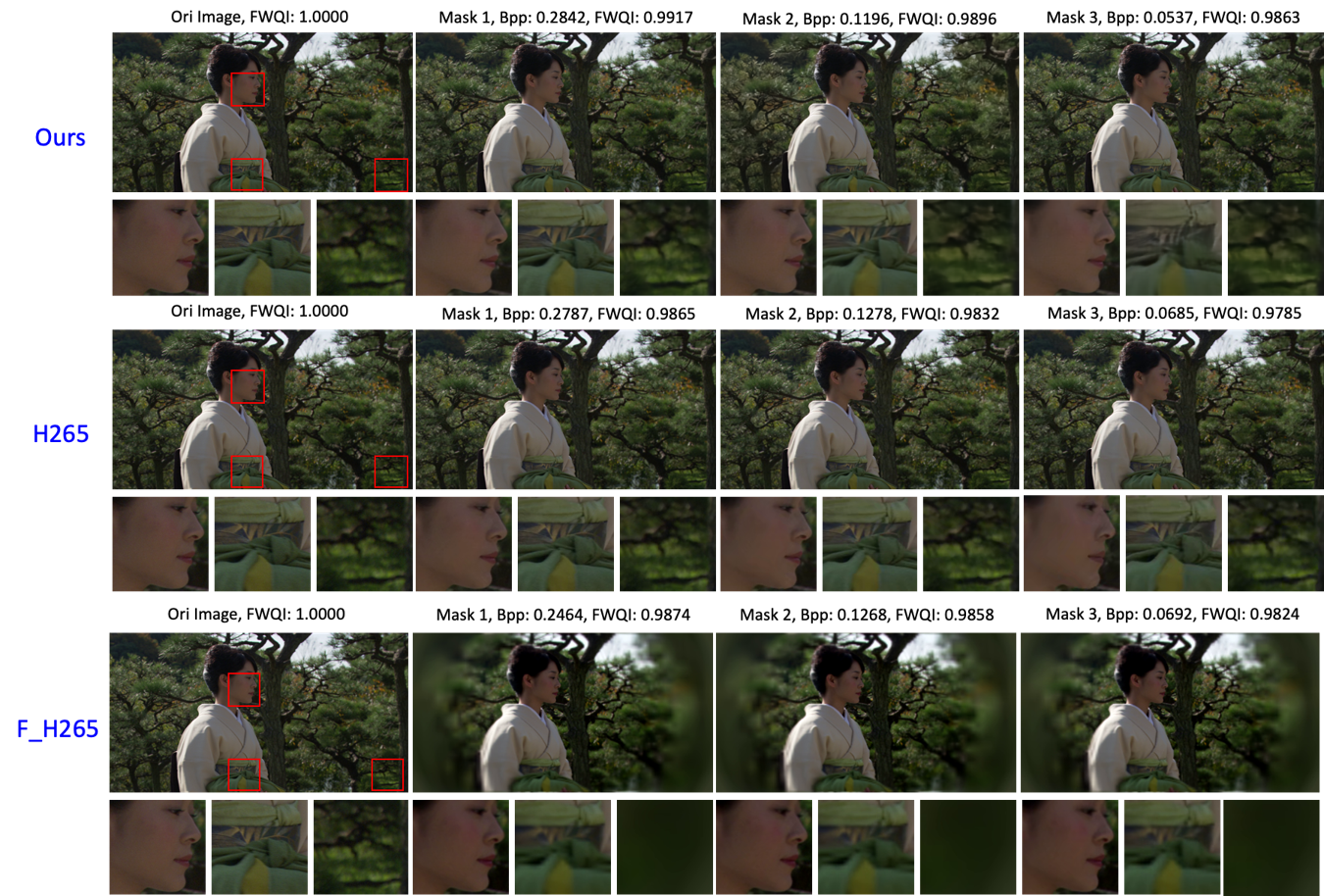}}
\caption{Examplar reconstructed foveated frames produced by FOV-MOVI-Codec, H.265, and Foveated H.265 (F\_H265), from top to bottiom.}
\label{fig:suppl_Visual}
\end{figure*}

\begin{figure*} [!h]
\centerline{
\subfigure[Reconstructed frame with $FMSC = H/2$]{
  \label{mask1}
\includegraphics[width=0.6\columnwidth]{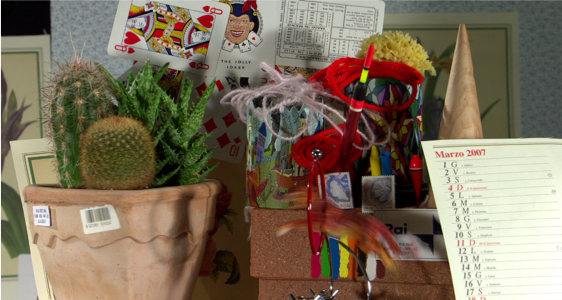}} 
\subfigure[Reconstructed frame with $FMSC = H/4$]{
  \label{mask2}
\includegraphics[width=0.6\columnwidth]{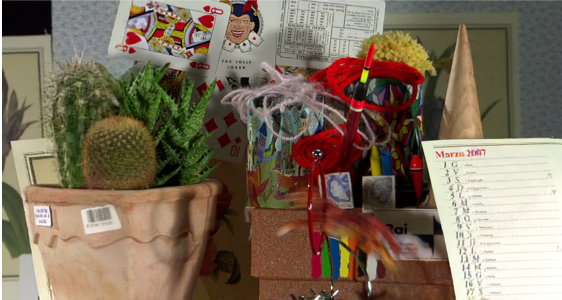}}
\subfigure[Reconstructed frame with $FMSC = H/6$]{
  \label{mask3}
\includegraphics[width=0.6\columnwidth]{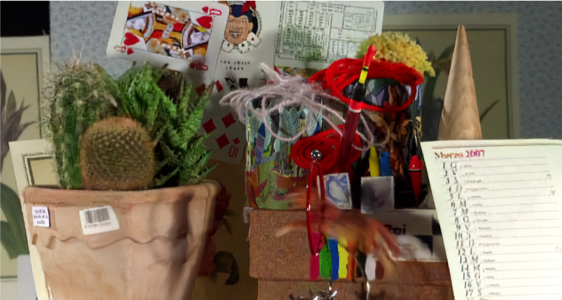}}
}
\centerline{
\subfigure[Difference image with $FMSC = H/2$]{
  \label{mask1}
\includegraphics[width=0.6\columnwidth]{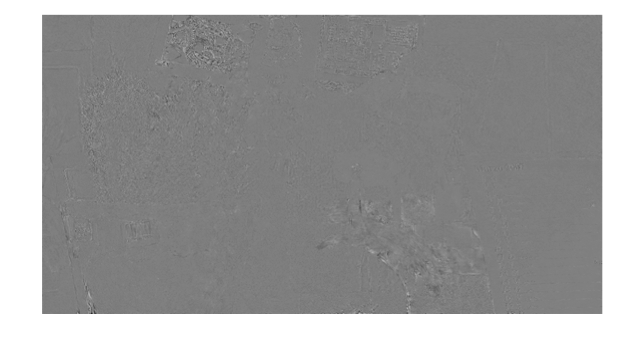}} 
\subfigure[[Difference image with $FMSC = H/4$]{
  \label{mask2}
\includegraphics[width=0.6\columnwidth]{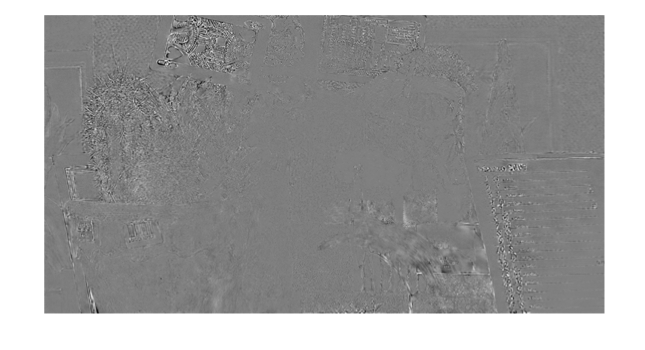}} 
\subfigure[[Difference image with $FMSC = H/6$]{
  \label{mask3}
\includegraphics[width=0.6\columnwidth]{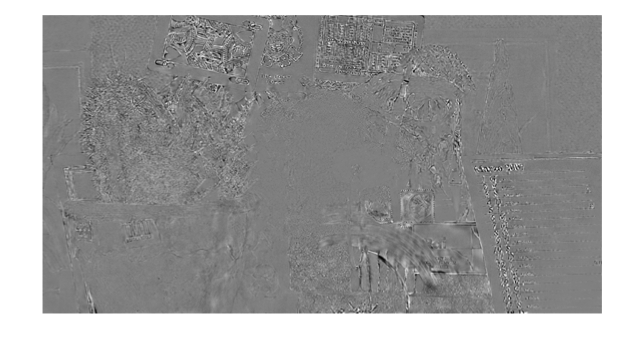}} 
}
\centerline{
\subfigure[Bits and SSIM Profile with FMSC$= H/2$]{
  \label{mask1}
\includegraphics[width=0.6\columnwidth]{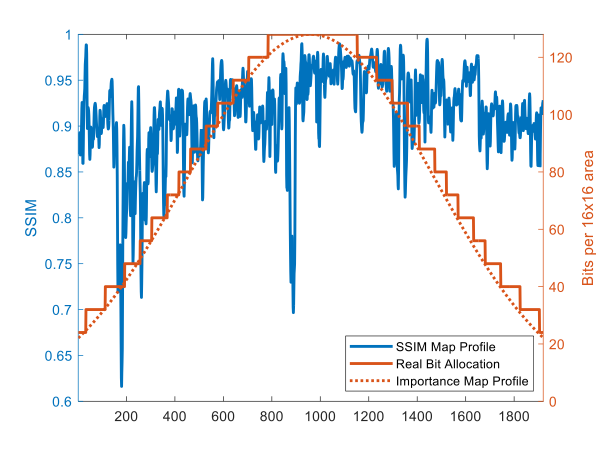}} 
\subfigure[Bits and SSIM Profile with FMSC$= H/4$]{
  \label{mask2}
\includegraphics[width=0.6\columnwidth]{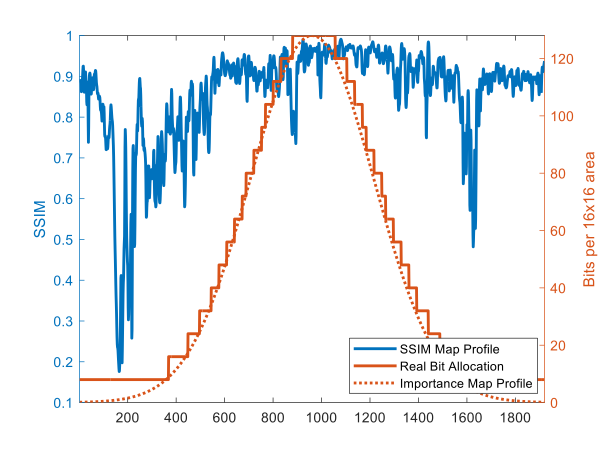}}
\subfigure[Bits and SSIM Profile with FMSC$= H/6$]{
  \label{mask3}
\includegraphics[width=0.6\columnwidth]{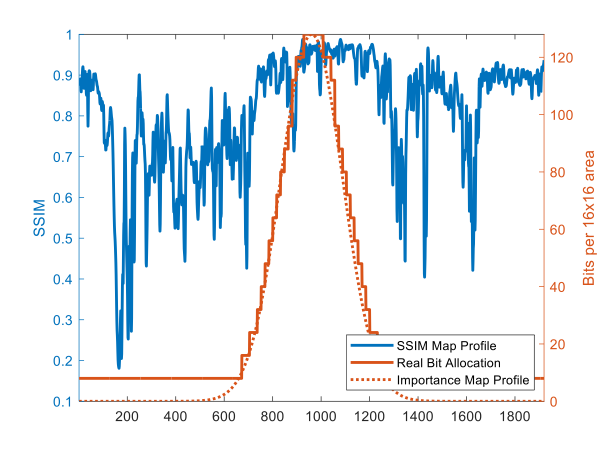}}
}
\caption{Reconstructed frames, differenced frames and bit-SSIM profiles for different foveation mask space constants (FMSCs).}
\label{fig:bits_2}
\end{figure*}

% that's all folks
\end{document}